\documentclass[prx,twocolumn,aps]{revtex4-1}
\usepackage{graphicx}
\usepackage{amsfonts}
\usepackage{amsmath,amssymb,amsthm,bm,multirow,longtable}
\usepackage{epsfig}
\usepackage{dcolumn}
\usepackage{epstopdf}
\usepackage{comment}
\usepackage{color}
\usepackage{hyperref}
\usepackage{bm}
\usepackage{float}

\DeclareMathOperator{\sech}{sech}

\begin{document} 

\title{The fastest pulses that implement dynamically corrected gates}

\author{Junkai Zeng}
\author{Edwin Barnes}%
 \email{efbarnes@vt.edu}
\affiliation{%
 Department of Physics, Virginia Tech, Blacksburg, Virginia 24061, USA
}%
\begin{abstract}
Dynamically correcting for unwanted interactions between a quantum system and its environment is vital to achieving the high-fidelity quantum control necessary for a broad range of quantum information technologies. In recent work, we uncovered the complete solution space of all possible driving fields that suppress transverse quasistatic noise errors while performing single-qubit operations. This solution space lives within a simple geometrical framework that makes it possible to obtain globally optimal pulses subject to a set of experimental constraints by solving certain geometrical optimization problems. In this work, we solve such a geometrical optimization problem to find the fastest possible pulses that implement single-qubit gates while cancelling transverse quasistatic noise to second order. Because the time-optimized pulses are not smooth, we provide a method based on our geometrical approach to obtain experimentally feasible smooth pulses that approximate the time-optimal ones with minimal loss in gate speed. We show that in the presence of realistic constraints on pulse rise times, our smooth pulses significantly outperform sequences based on ideal pulse shapes, highlighting the benefits of building experimental waveform constraints directly into dynamically corrected gate designs.
\end{abstract}

\maketitle

\section{Introduction}

The principles of quantum mechanics enable novel technologies such as quantum sensing, simulation, and computing \cite{Nielsen_Chuang,degen2017quantum,georgescu2014quantum}. One of the main obstacles in realizing such technologies is the decoherence caused by unwanted interactions between a quantum system and its environment \cite{chirolli2008decoherence,Bialczak_PRL07}. While quantum error correction codes have the potential to alleviate this problem, it remains important to improve control fidelities as much as possible to surpass error correction thresholds and to lessen the demanding resource requirements of such schemes\cite{gottesman2010introduction,preskill1998reliable,aliferis2007level,fowler2012surface}.

The Hahn spin echo protocol first introduced in the context of nuclear magnetic resonance \cite{Hahn_PR50} gradually inspired the broader concept of dynamical decoupling\cite{viola1999dynamical}, or more generally, dynamically corrected gates (DCGs)\cite{khodjasteh2009dynamically}, where deviations in the evolution of a quantum system caused by noise fluctuations or ensemble inhomogeneities in the system Hamiltonian are dynamically suppressed by applying carefully designed driving pulses\cite{Carr_Purcell,Meiboom_Gill,Viola_PRA98,Uhrig_PRL07,yang2008universality,khodjasteh2005fault,Wang_NatComm12,Wang_PRA14,Merrill_Wiley14,Kestner_PRL13,khodjasteh2012automated,khodjasteh2009dynamically,khodjasteh2009dynamical,khodjasteh2010arbitrarily,jones2012dynamical,green2012high,green2013arbitrary}. The goal of dynamical decoupling is to remove such errors in order to preserve the state of the system, while DCGs constitute a broader class of control schemes that also implement nontrivial operations while suppressing errors. Many DCG schemes are constructed by designing two or more noisy quantum operations such that their combined evolution yields the desired gate and such that their errors cancel to at least first order\cite{Wang_NatComm12,Wang_PRA14,Merrill_Wiley14,Kestner_PRL13,khodjasteh2012automated,khodjasteh2009dynamically,khodjasteh2009dynamical,khodjasteh2010arbitrarily}. This provides a general approach, however arranging for the error cancellation often requires restricting to a fixed set of pulse shapes and solving highly nonlinear algebraic equations. This can make it challenging to control the length of the resulting pulse sequence, and hence the duration of the gate. On the other hand, devising a more direct method to ensure the cancellation of errors is made difficult by the fact that it is hard to solve the time-dependent Schrodinger equation analytically aside from a few special pulse shapes. Over the last few years, progress in this direction has been made by using reverse-engineering techniques to solve the Schrodinger equation\cite{fanchini2007continuously,Barnes_PRL12,Barnes_SciRep15,Barnes_PRA13}.

In a recent work\cite{zeng2018general}, we introduced a new method that enables one to obtain all possible smooth driving pulses that perform dynamical gate correction in the case of a resonantly driven qubit subject to transverse quasistatic noise. This method exploits the remarkable finding that the evolution of a single qubit state can be mapped onto a curve in a 2D plane such that the driving pulse amplitude and evolution time correspond to the the curvature and length along the curve, respectively. The fact that the complete solution space of DCG pulses lives within this simple geometrical framework opens up the possibility of not only cancelling noise but of finding the globally optimal pulses that do so given a specific set of experimental constraints on the pulse shape. This is in contrast to numerical optimal control theory techniques\cite{palao2002quantum,rabitz2000whither,brif2010control}, which typically yield only local extrema of a chosen cost function, and often produce complicated pulse shapes. 

One of the main criteria used to benchmark the quality of a qubit platform is the number of gate operations that can be performed within the coherence time of the qubit. Thus, tremendous effort has been devoted to extending coherence times as much as possible in a variety of physical systems\cite{muhonen2014storing,martins2016noise,paik2011observation,wang2017single}. Many works have also focused on improving this number by speeding up gate times using carefully tailored control schemes \cite{carlini2006time,wang2015quantum,chen2015near}. However, the question of how to reduce gate times while performing dynamical gate correction has remained an open problem.

In this paper, we show how to find the fastest possible pulses that implement a desired qubit operation while canceling noise to second order. We solve this problem by leveraging our geometrical framework to recast gate time minimization as a geometrical optimization problem that can be solved using variational calculus. We find that the time-optimal pulses lie within a certain class of composite square pulses. To avoid the difficulties in realizing square pulses, we show that such pulses can be approximated with smooth pulses using a systematic procedure based on our geometrical framework that ensures that the pulses remain fast and the errors continue to be strongly suppressed. We show that our smooth, time-optimized pulses signficantly outperform square-pulse DCG sequences when experimental rise time constraints are taken into account. 

The paper is organized as follows. In Sec.~\ref{sec:ham}, we introduce the qubit Hamiltonian and noise model and briefly review the geometrical framework. In Sec.~\ref{sec:time}, we cast the gate time optimization problem as a geometrical optimization problem and solve it using variational calculus. In Sec.~\ref{sec:smooth}, we present a systematic algorithm for smoothing the time-optimal pulses and compare the results with naive implementations of square pulse sequences. We give some concluding remarks in Sec.~\ref{sec:conclusions}. Two appendices provide further details of the geometrical formalism and of our derivation of the time-optimal pulses.

\section{Hamiltonian and geometric framework}\label{sec:ham}

A widely used semi-classical model describing the decoherence of a single qubit subject to slow noise is given by the following Hamiltonian:
\begin{equation}
\mathcal{H}(t)=\frac{\Omega(t)}{2}\sigma_z+\delta\beta\sigma_x.
\label{eq:hamil}
\end{equation}
Here $\sigma_z$ and $\sigma_x$ are Pauli matrices, $\Omega(t)$ is the driving field (or driving field amplitude if we view $\mathcal{H}(t)$ as the Hamiltonian for a resonantly driven qubit in the rotating frame defined by the driving), and the off-diagonal term $\delta\beta$ is a stochastic variable describing transverse noise. We work in the quasistatic limit where $\delta\beta$ is viewed as an unknown quantity that remains constant during a single gate operation but which can vary between different runs of an experiment. Decoherence results from averaging measurements over many such runs. Designing DCGs amounts to finding choices of $\Omega(t)$ that produce a desired evolution operator $\mathcal{U}(T)$ (at some final time $T$) that is insensitive to $\delta\beta$. Due to the fact that $\mathcal{H}(t)$ doesn't commute with itself at different times, solving the Schr\"odinger equation to obtain $\mathcal{U}(t)$ in closed form is hard, and without such an expression it is difficult to ensure that $\mathcal{U}(T)$ is insensitive to the noise. (Note that the reverse-engineering method introduced in \cite{Barnes_PRL12,Barnes_SciRep15} does not apply for a Hamiltonian of the form of Eq.~\eqref{eq:hamil}.) 

In Ref.~\cite{zeng2018general}, we uncovered a surprising geometrical structure hidden within the evolution operator $\mathcal{U}(t)$ generated by a Hamiltonian of the form of \eqref{eq:hamil} that allows us to avoid the difficulties with solving the Schr\"odinger equation. In particular, we showed that the evolution can be mapped to a curve living in a 2D plane. This curve is obtained by expanding $\mathcal{U}(t)$ to first order in $\delta\beta$; the time-dependent complex coefficient of the first-order term can be thought of as tracing out a curve in the 2D complex plane as time progresses. The distance from the origin of the plane to a point on the curve thus provides a measure of the first-order error in $\mathcal{U}(t)$ at a particular moment in time. We review this geometrical framework in Appendix~\ref{app:derive}. Two of its key features are that the evolution time is measured by the distance (arc length) along the curve:
\begin{equation}
	t(\lambda)=\int_0^\lambda \sqrt{x'(\mu)^2+y'(\mu)^2}d \mu,
	\label{eq:time}
\end{equation}
and the driving field at each time is equal to the curvature, $\kappa(\lambda)$, at the corresponding point on the curve:
\begin{equation}
	\kappa(\lambda)=\Omega(t(\lambda))=\frac{x'y''-y'x''}{(x'^2+y'^2)^{3/2}}.
	\label{eq:kappa}
\end{equation}
Here, we specify a curve in terms of the parametric functions $x(\lambda)$ and $y(\lambda)$, which are the real and imaginary parts of the first-order error coefficient in $\mathcal{U}(t)$, and $\lambda$ is a parameter along the curve. From the definition of the curve, it is clear that we must require it to form a closed loop, $x(\lambda(T))=y(\lambda(T))=0$, in order to guarantee that the first-order error in $\mathcal{U}(T)$ vanishes. In this case, the operation undergone by the qubit is a $z$-rotation by angle $\phi+\pi$, where $\phi$ is the angle subtended by the two legs of the curve at the origin. An example is shown in Fig. \ref{fig:curveexample}. As we also review in Appendix~\ref{app:derive}, canceling the second-order error additionally requires the net area encosed by the curve to vanish.
\begin{figure}
 \centering
\includegraphics[width=0.6\columnwidth]{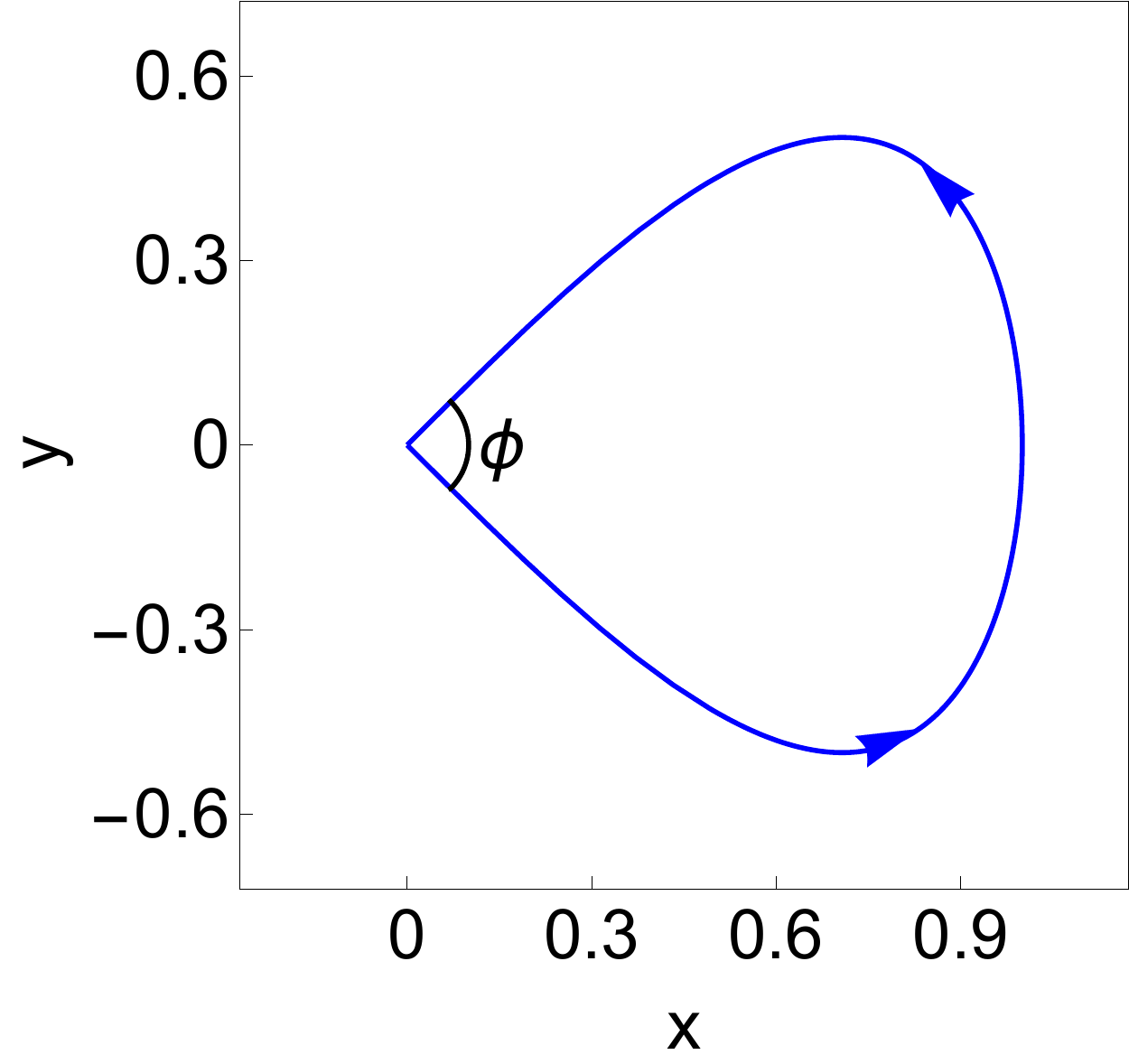}
\caption{An example of a closed curve, with angle subtended by the two legs $\phi$.}
\label{fig:curveexample}
\end{figure}

It is important to emphasize that these geometric error-cancellation constraints are both necessary and sufficient and thus provide the complete solution space of DCGs for Eq.~\eqref{eq:hamil}. Any curve $\{x(\lambda),y(\lambda)\}$ satisfying these constraints can be mapped to a driving pulse $\Omega(t)$ that performs dynamical gate correction simply be extracting the curvature of the curve. In addition to recovering standard $\delta$-pulse or square-pulse DCG schemes, this framework also allows us to generate smooth driving pulses simply by choosing curves that are smooth to second order. It is clear from this construction that the DCG solution space is vast and offers infinitely many solutions for every desired gate operation, leaving plenty of room for further pulse optimization depending on the capabilities and requirements of a given experiment.

\section{Gate time optimization}\label{sec:time}

Our task here is to find the fastest driving pulses that implement a desired gate operation while suppressing the effects of noise. If we do not take into account realistic experimental constraints on the pulse shape, then this question does not have a well defined answer since $z$ gates generated by Eq.~\eqref{eq:hamil} can be made arbitrarily fast simply by making the pulse amplitude (equivalently the driving power) arbitrarily large. From the point of view of the geometrical framework, the latter amounts to shrinking the overall scale of the curve, which results in larger curvature. Of course, in any real physical qubit system, there will be a limit on the driving power that can be applied, and we take this into account by placing a restriction on the pulse amplitude:
\begin{equation}
	|\Omega(t)|\le1.\label{eq:ampconstraint}
\end{equation}
Here, we have chosen the upper bound to be unity for simplicity and without loss of generality; this amounts to choosing the maximal pulse amplitude as our basic energy unit. We can reinterpret the results that follow in terms of any maximal pulse amplitude by rescaling the pulse amplitude and duration in such a way that the product $T\Omega(t)$ remains invariant. 

Geometrically, Eq.~\eqref{eq:ampconstraint} imposes an upper bound on the curvature of the plane curve. The problem of looking for the fastest pulses then amounts to searching for the shortest curves that respect this curvature bound while also satisfying error-cancellation constraints (closed curve, zero net area) and the target rotation constraint (angle subtended at origin). For now, we focus on cancelling the first-order noise as a starting point, and then turn to also cancelling the second-order noise errors afterward.

It is straightforward to express this geometrical plane curve optimization problem as a variational calculus problem, where the action is given by
\begin{equation}
\begin{split}
&S[y(\lambda),x(\lambda)]=\\
&\int_0^{\lambda(T)}\!\!\!d\mu\left[\sqrt{x'(\mu)^2+y'(\mu)^2}+
\alpha(\mu)\left(\kappa(\mu)-\tanh(\omega(\mu))\right)\right].
\end{split}
 \end{equation}
The first term is the length of the curve, while the second term imposes the upper bound on the curvature $\kappa(\mu)$, which is given in Eq.~\eqref{eq:kappa}. We have introduced both a Lagrange multiplier, $\alpha(\mu)$, and a slack variable, $\omega(\mu)$, since Eq~\eqref{eq:ampconstraint} is an inequality constraint. We need to minimize the action while respecting the boundary conditions
$x(0)=x(\lambda(T))=y(0)=y(\lambda(T))=0$ and $\arctan\left(y'/x'\right)|_0^{\lambda(T)}=\phi+\pi$, which ensure that the first-order error cancels and that the qubit rotation angle is $\phi+\pi$, respectively. 

Varying the action with respect to $\omega$ gives
\begin{equation}
	\alpha \sech(\omega)\tanh(\omega)=0,
	\label{eq:alphasechtanh}
\end{equation}
while the variation with respect to $\alpha$ yields 
\begin{equation}
	\kappa= \tanh(\omega).
	\label{eq:curvature}
\end{equation}
The solutions to Eq. \eqref{eq:alphasechtanh} are $\omega=0,\pm\infty$, and plugging these into Eq. \eqref{eq:curvature} gives curvature $\kappa= 0,\pm1$. The first solution corresponds to the case where the curve becomes a straight line, and the latter ($\kappa=\pm1$) yields circular arcs with radius $r=1$. Therefore, any curve comprised solely of straight lines and $r=1$ arcs is a local time-optimal solution. One requirement worth mentioning is that the straight lines and arcs must connect to each other tangentially, as kinks or cusps give rise to infinite curvature and thus violate Eq.~\eqref{eq:ampconstraint}.

\begin{figure}
 \centering
\includegraphics[width=0.6\columnwidth]{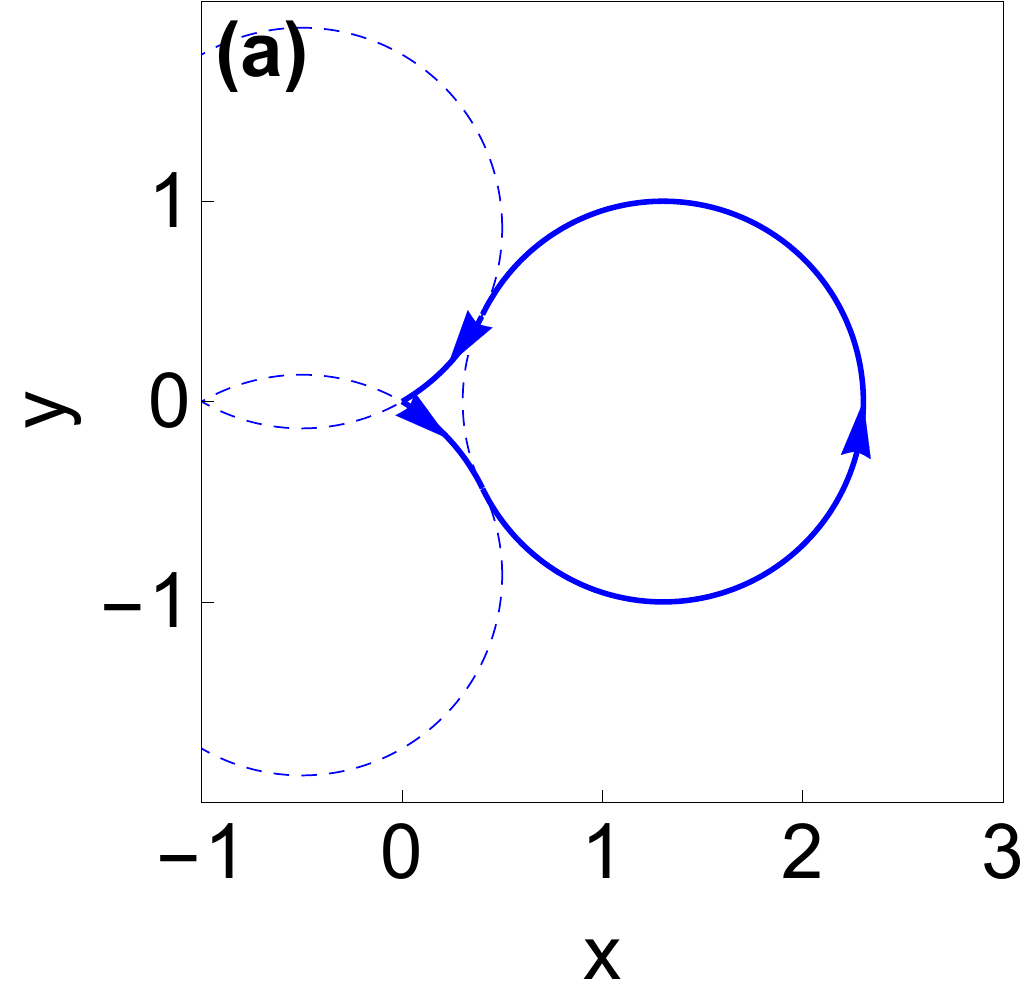}
\includegraphics[width=0.8\columnwidth]{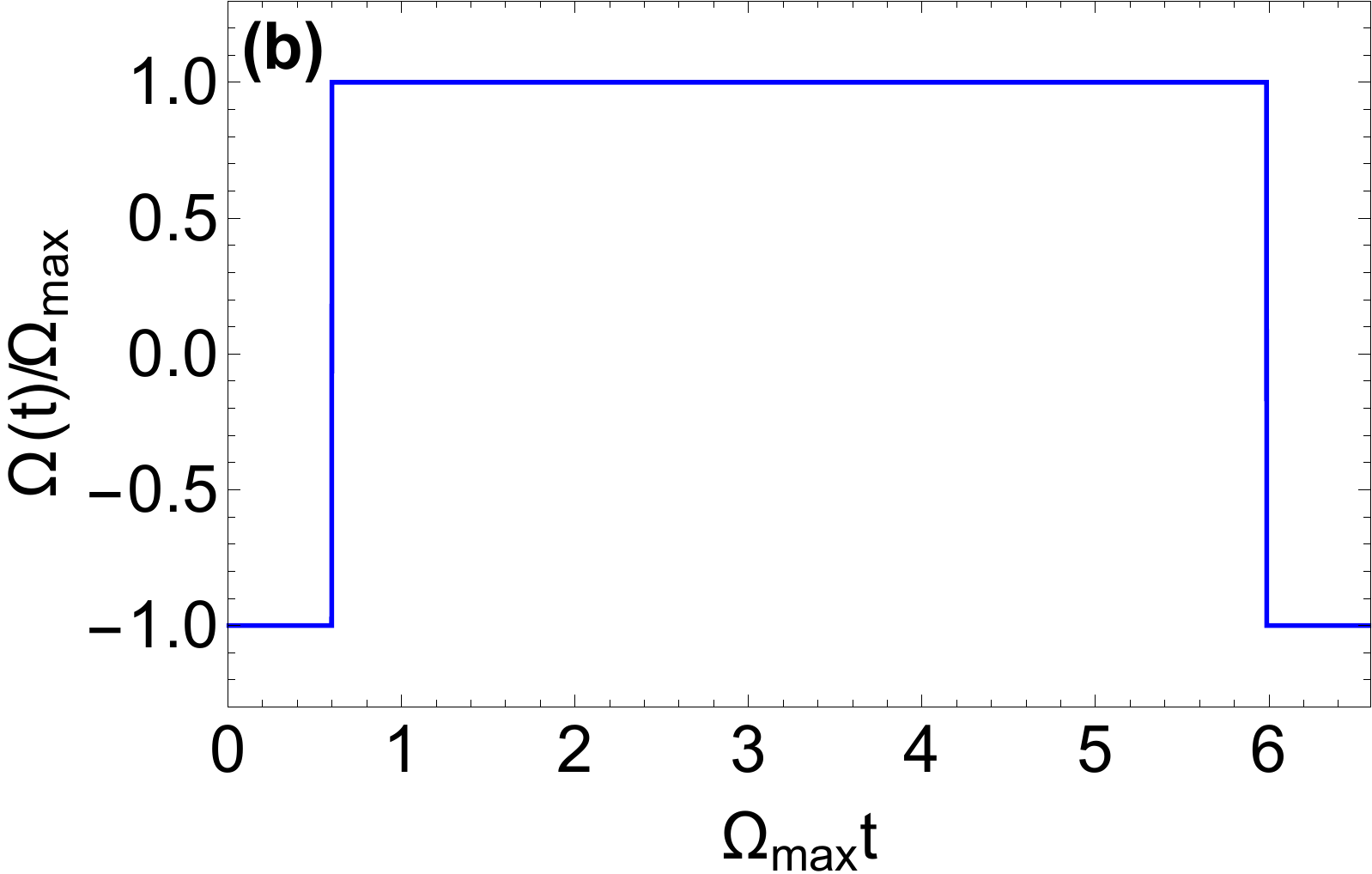}
\caption{(a) Curve that yields the fastest qubit $z$-rotation by angle $4\pi/3$ (the angle subtended at the origin is $\phi=\pi/3$) while canceling the first-order noise error and respecting the pulse constraint $|\Omega(t)|\le\Omega_{max}$. The curve is made up of three circular arcs that connect tangentially to each other. (b) Composite square pulse obtained from the curve in (a). The minimal gate time in this case is $T_{min}=6.59/\Omega_{max}$.}
\label{fig:firstorder}
\end{figure}

\begin{figure}
 \centering
\includegraphics[width=0.8\columnwidth]{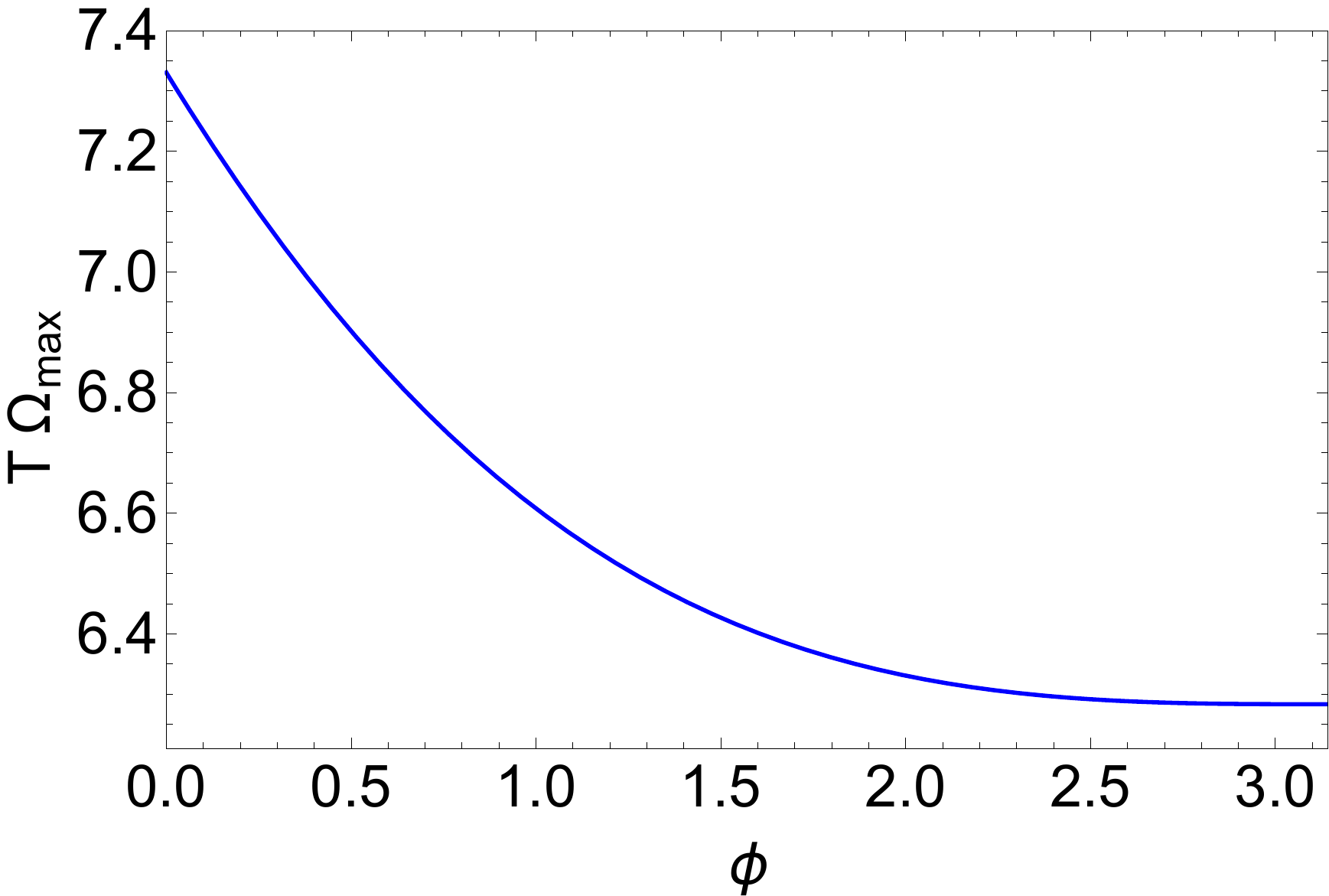}
\caption{The minimal time to perform quantum rotations while canceling first-order noise for different values of $\phi$.}
\label{fig:length1plot}
\end{figure}

The solution above tells us that time-optimal pulses map to curves made up of straight lines and circular arcs, but it does not tell us how to assemble these components to obtain the absolute fastest pulse. However, given that we have very few components to work with, this task can be completed through a simple process of elimination. We first focus on the shape of the curve near the origin of the 2D plane, where we know that the curve must form a cusp with opening angle $\phi$ in order to achieve the desired target $z$-rotation by angle $\phi+\pi$. This cusp must be formed from either two intersecting straight lines or from two intersecting $r=1$ circular arcs. The cusp could also be formed from a straight line intersecting a circular arc, but here we assume without rigorous proof that the optimal solutions are mirror symmetric about an axis which we take to be $x$ without loss of generality (any two curves related by a rigid rotation about the origin are equivalent in the sense that they yield the same driving field \cite{zeng2018general}). In the case where the cusp is formed by two straight lines, there is only one way to complete a smooth closed loop using only one additional segment (a circular arc). On the other hand, there are two ways to form a smooth loop by inserting one additional segment (either a straight line or a circular arc) in the case where the cusp at the origin is formed from two arcs. These resulting curves are described and compared in detail in Appendix~\ref{app:construction}, where we show that the shortest path overall is obtained by using only circular arcs and no straight lines at all. An example of such curve, where the subtended angle is $\phi = \frac{\pi}{3}$, is shown in Fig.\ref{fig:firstorder}(a). Since circular arcs have constant curvature, they correspond to driving pulses of constant amplitude, i.e., square pulses. We thus find that the global time-optimal pulse is a composite square pulse. For a rotation angle of $\phi+\pi$, the explicit form of this composite pulse is
\begin{equation}
	\begin{split}
	\Omega(t)&=\begin{cases}
	-1 & 0<t<\psi-\phi/2\\
	1 & \psi-\phi/2<t<3\psi-\phi/2+\pi\\
	-1 & 3\psi-\phi/2+\pi<t<4\psi-\phi+\pi
	\end{cases},\\
	\psi&=\arccos(\tfrac{1}{2}\cos(\phi/2)).
	\end{split}
\end{equation}
This pulse is depicted in Fig.\ref{fig:firstorder}(b) for the case $\phi=\pi/3$. We see from this result that the absolute minimal time to perform a $z$-rotation by angle $\phi+\pi$ while canceling the first-order noise is $T_{min}=[4\arccos(\tfrac{1}{2}\cos(\phi/2))-\phi+\pi]/\Omega_{max}$. A plot of this function is shown in Fig.\ref{fig:length1plot}. We can see that the shortest quantum rotation is the identity operation with $\phi=\pi$. A closed loop in this case should contain an internal circle, and the $\phi=\pi$ case is where the curve itself becomes an exact perfect circle(no cusp at the origin). As $\phi$ becomes smaller, extra segments in the curve are necessary to complete the loop, which causes the minimal time to increases as the rotation angle decreases.

We now extend this analysis to cancel the second-order noise errors as well. Since the second-order error is proportional to the area enclosed by the curve, we need to add one more term in the action to enforce the condition that the enclosed area vanishes. Here to simplify the math, we will again assume that the curve is symmetric about the $x$-axis, and instead of optimizing with respect to two functions, $x(\mu)$ and $y(\mu)$, in the action, we use one function, $y(x)$. In this case, only half of the curve needs to be considered. The optimization problem can then be formulated using the following action:
\begin{equation}
\begin{split}
&S[y(x)]=\\
&\int_0^{x_0}\!\!\!\!dx\left[\sqrt{1+y'(x)^2}+\alpha(x)\left(\tilde{\kappa}(x)-\tanh(\omega(x))\right)+\Gamma y(x)\right],
\end{split}
\end{equation}
where $\Gamma$ is a Lagrange multiplier imposing the vanishing area constraint, and $\tilde{\kappa}(x)=y''(x)/[1+y'(x)^2]^{3/2}$ is the curvature. We must also impose the boundary conditions $y(0)=y(x_0)=0$, $y'(0)=-\frac{ \phi}{2}$ and $y'(x_0)=\pm \frac{\pi}{2}$ to ensure that the two halves of the curve join smoothly together at $x=x_0$ and form a cusp of angle $\phi$ at $x=0$. Eq.~\eqref{eq:alphasechtanh} still holds, but now the Euler-Lagrange equation for $y(x)$ in the $\alpha=0$ case yields a new type of curve segment when $\Gamma\ne0$. This equation reads
\begin{equation}
   y''(x)=\Gamma\left[1+y'(x)^2\right]^{\frac{3}{2}},
\end{equation}
and has the solutions 
\begin{equation}
	(y-C_1)^2+(x-C_2)^2=1/\Gamma^2,
\end{equation}
for arbitrary constants $C_1$ and $C_2$. This corresponds to a circular arc with radius $r=1/\Gamma$ centered at $\{x,y\}=\{C_1,C_2\}$. When $\Gamma=0$, this additional solution reduces to a straight a line, and we are left with only straight lines and $r=1$ arcs (which again come from solutions with $\alpha\ne0$, $\omega=\pm\infty$) as we found in the case of first-order error cancellation. Returning to $\Gamma\ne0$, we have found that a curve corresponding to locally time-optimal, second-order robust driving pulses is made up of two types of circular arcs, one with $r=1$ and the other with $r=1/\Gamma$, and straight lines. Note that we must also impose $\Gamma\le1$ for consistency with the pulse amplitude constraint, Eq.~\eqref{eq:ampconstraint}. The resulting optimal solutions are once again composite square pulses.

\begin{figure}
 \centering
\includegraphics[width=0.6\columnwidth]{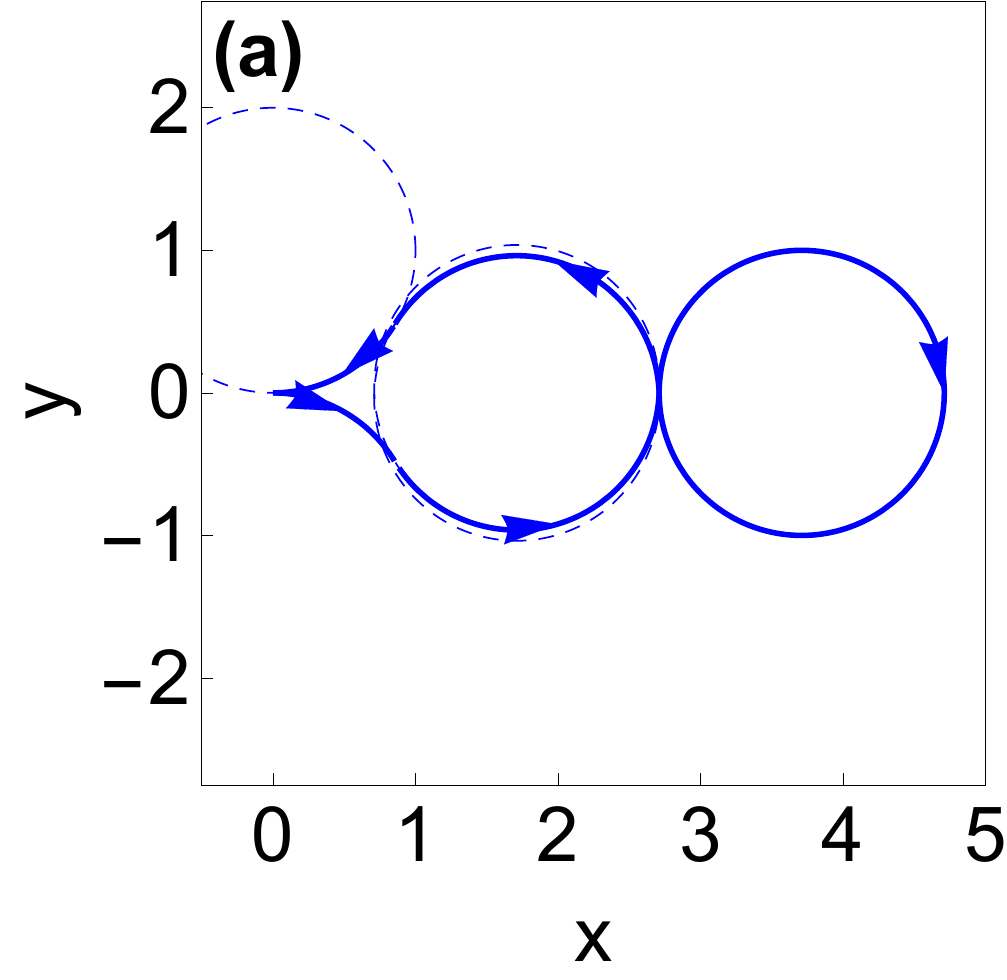}
\includegraphics[width=0.8\columnwidth]{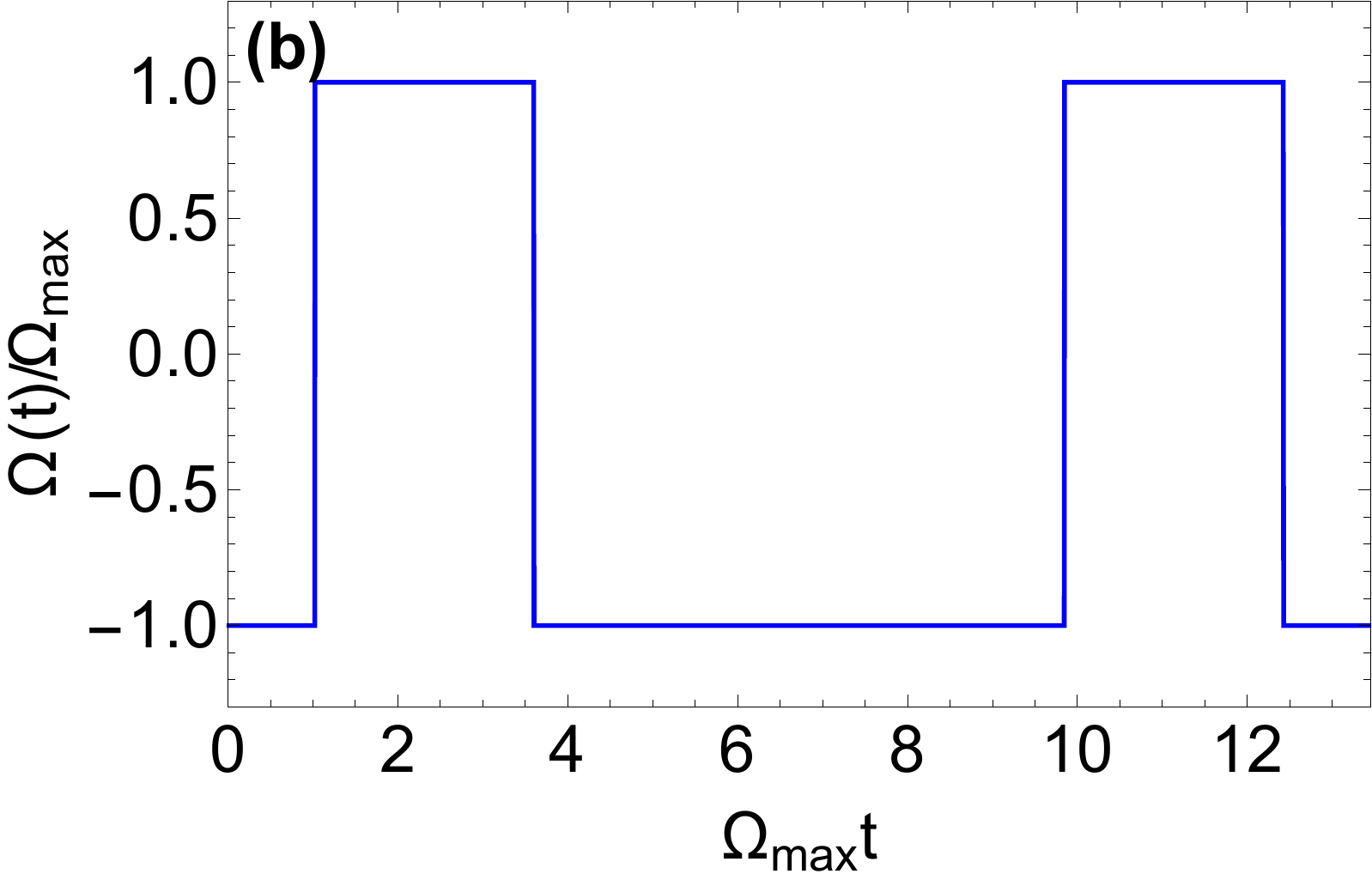}
\includegraphics[width=0.8\columnwidth]{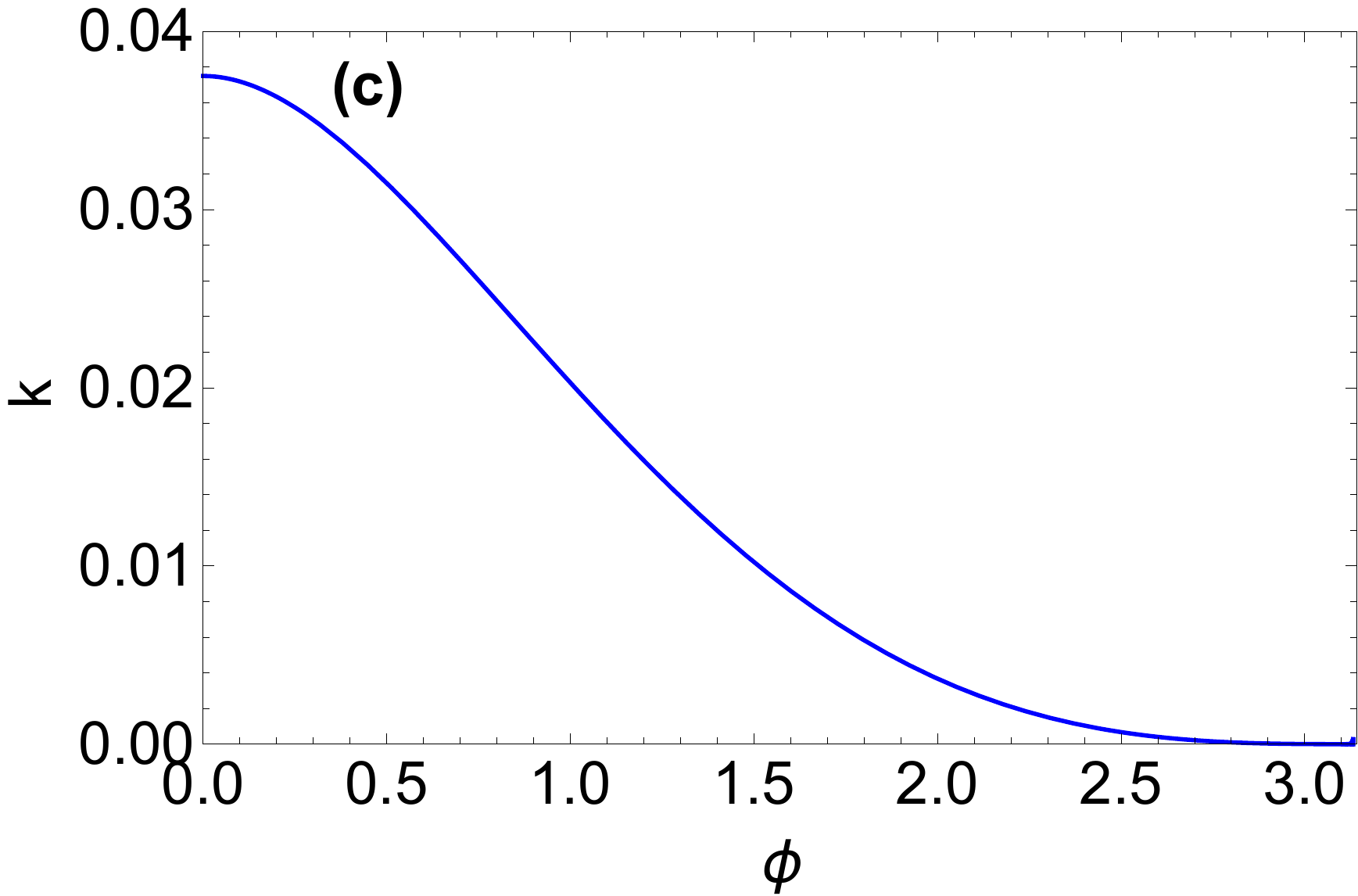}
\caption{(a) Curve that yields the fastest qubit $\pi$-rotation (subtended angle $\phi=0$) while canceling the second-order noise error and respecting the pulse constraint $|\Omega(t)|\le\Omega_{max}$. (b) Corresponding driving pulse. (c) Parameter $k(\phi)$ as a function of the subtended angle $\phi$.}
\label{fig:secondorder}
\end{figure}

\begin{figure}
 \centering
\includegraphics[width=0.8\columnwidth]{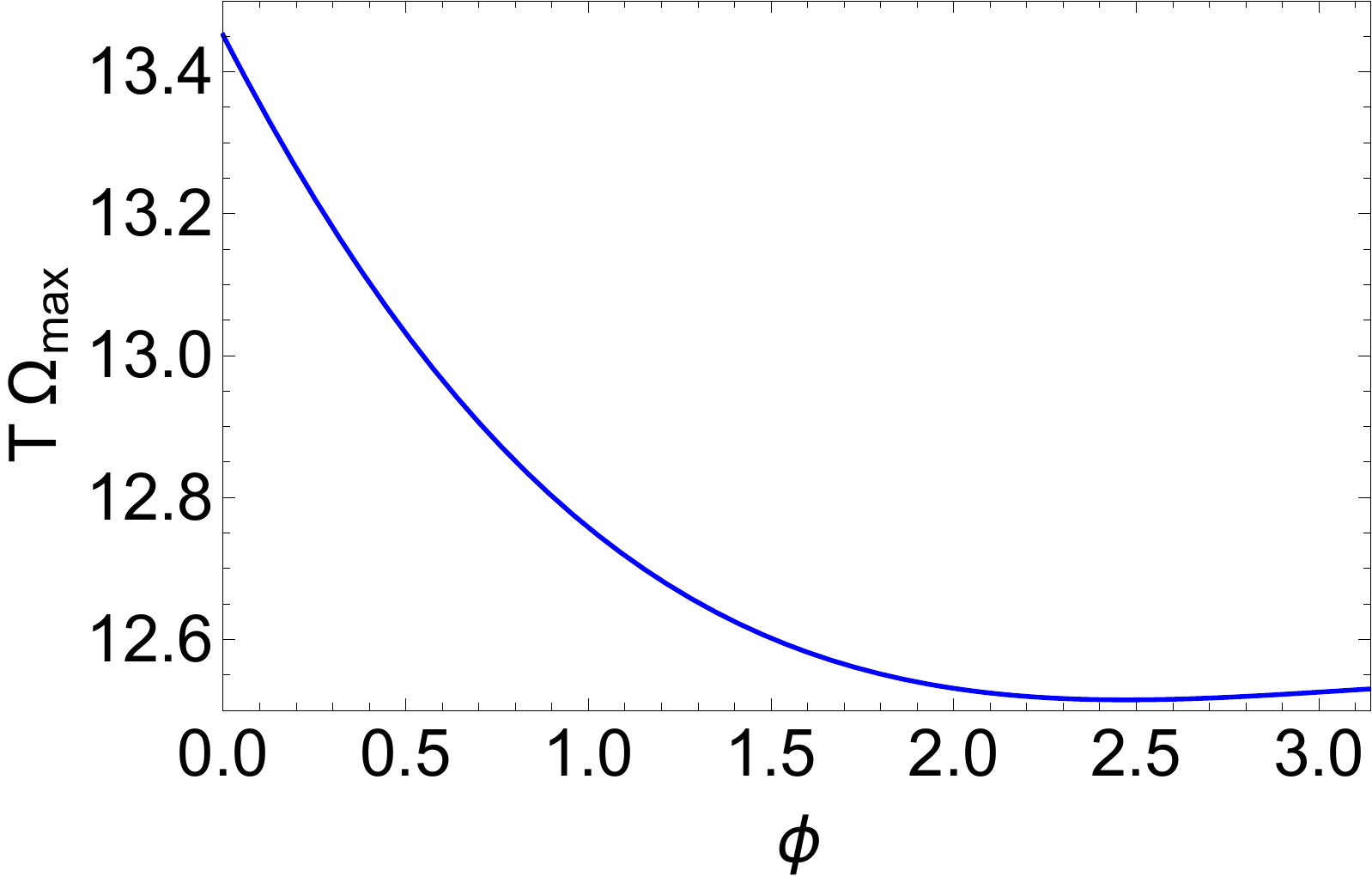}
\caption{The minimal time to perform quantum rotations while canceling second-order noise for different values of $\phi$.}
\label{fig:secondorderlength}
\end{figure}

By comparing several possible arrangements, we found that similar to the first-order case, the shortest possible curve satisfying all these requirements is still comprised solely of connected $r=1$ arcs. Including straight lines or $r=1/\Gamma$ arcs generally slows down the gate. A detailed discussion of how one can systematically construct curves made from $r=1$ arcs that obey all the constraints is given in Appendix~\ref{app:construction}. An example curve with $\phi=0$ (corresponding to a $\pi$ rotation) and its corresponding pulse are shown in Figs.~\ref{fig:secondorder}(a),(b). For a general rotation angle, five such arcs are needed, and the driving pulse $\Omega(t)$ is given by
\begin{equation}
\begin{split}
	\Omega(t)&=
	\begin{cases}
	-1 & t<\psi_1-\frac{\phi}{2}\\
	1 & \psi_1-\frac{\phi}{2}<t<2\psi_1+\psi_2-\frac{\phi}{2}\\
	-1 & 2\psi_1+\psi_2-\frac{\phi}{2}<t<2\psi_1+3\psi_2-\frac{\phi}{2}+\pi\\
	1 & 2\psi_1+3\psi_2-\frac{\phi}{2}+\pi<t<3\psi_1+4\psi_2-\frac{\phi}{2}+\pi\\
	-1 & 3\psi_1+4\psi_2-\frac{\phi}{2}+\pi<t<4\psi_1+4\psi_2-\phi+\pi\\
	\end{cases},\\\label{eq:2ndorderpulse}
	\psi_1&=\arccos( \frac{k(\phi)+\cos(\frac{\phi}{2})}{2}),\\
	\psi_2&=\arccos(\frac{k(\phi)}{2}).
\end{split}
\end{equation}
Here $k(\phi)$ is a parameter which measures the vertical distance from the $x$-axis to the centers of the two middle $r=1$ arcs in Fig.~\ref{fig:secondorder}(a). This distance is determined numerically by enforcing the zero-net-area constraint. A plot of the resulting $k(\phi)$ is shown in Fig.\ref{fig:secondorder}(c). We see that as $\phi\to\pi$ (corresponding to an identity operation), the cusp at the origin gradually disappears, and the curve becomes two tangential circles of equal area but opposite orientation. For a general $\phi$, we see that the absolute minimal time to perform the rotation while canceling second-order noise is $T_{min}=[4\psi_1+4\psi_2-\phi+\pi]/\Omega_{max}$. A plot of this function is shown in Fig.\ref{fig:secondorderlength}, where it is again evident that larger values of $\phi$ lead to faster gates.

\section{Pulse smoothing protocol}\label{sec:smooth}

We have seen that the fastest pulses which cancel the leading-order noise errors are composite square pulses. Although these pulses respect the experimental constraint of finite pulse amplitude, they are not yet experimentally practical because they require infinitely fast pulse rise times. However, they still constitute a good starting point for designing smooth pulses that accomplish the same tasks at speeds close to the global minimum. A naive approach to obtaining such pulses would be to directly construct a smooth interpolation of the optimal composite square pulses. For example, we could replace each abrupt step from $\Omega=-1$ to $\Omega=1$ in the sequence by a smooth interpolation function $f(t)=\tanh(\mathcal{R}(t-t_0))$, where $\mathcal{R}$ is a parameter controlling the rise time. However, with such an approach, there is no guarantee that the error-cancellation constraints will continue to be even approximately satisfied as $\mathcal{R}$ is reduced to satisfy a particular rise time constraint. A more sophisticated approach that instead utilizes Fourier analysis and Chebyshev polynomials to achieve this exists \cite{jones2012dynamical}, but this technique has only been applied in the case of identity operations so far and has not yet been extended to single-qubit rotations.

\begin{figure}
\centering
\includegraphics[width=\columnwidth]{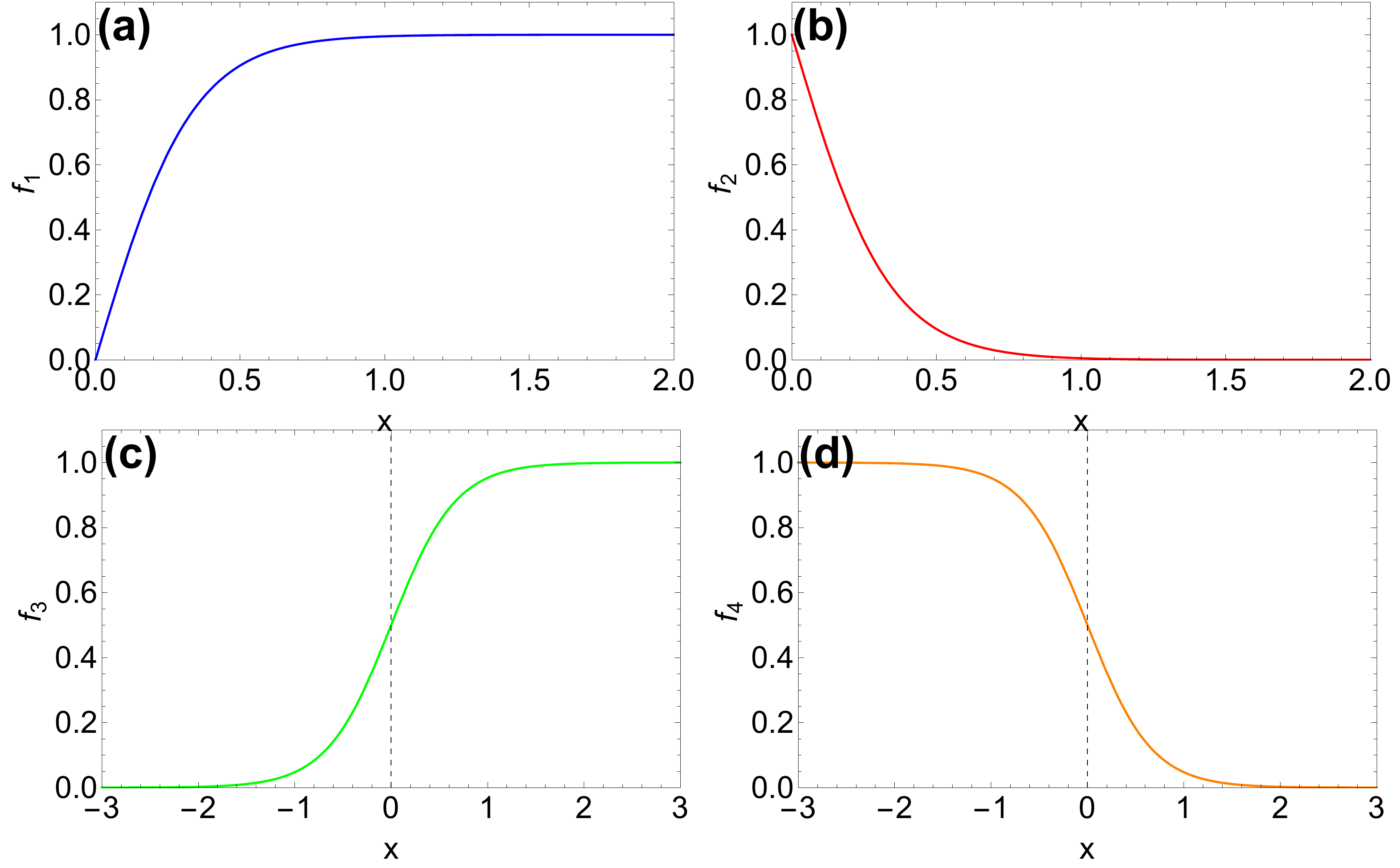}
\caption{Toolbox for smoothing the curve. (a) $f_1(\mathcal{P}, x)=\tanh(\mathcal{P} x)$. (b) $f_2(\mathcal{P}, x)=1-f_1(\mathcal{P}, x)$. Together with $f_1$, we use this to guarantee that the second derivative of the final smoothed function (and hence the driving pulse) starts from zero and ends at zero. (c) $f_3(\mathcal{Q}, x)=\left[1+\exp(-\mathcal{Q} x)\right]^{-1}$. (d) $f_4(\mathcal{Q}, x)=1-f_3(\mathcal{Q}, x)$. Together with function $f_3$, we use this to smoothly connect different pieces of the curve such that the end result is infinitely differentiable everywhere.}
\label{fig:smoothing}
\end{figure}

We can solve this problem by taking advantage of the geometrical framework, which provides the option to perform the smoothing on the plane curve instead of smoothing the pulse directly. This allows us to ensure that the error-cancellation constraints are still approximately satisfied after the smoothing has been performed. The basic strategy we follow is to connect the piecewise functions for $x(\lambda)$ and $y(\lambda)$ in such a way that they both become fully analytical at every point $\lambda$. To do this, we make use of four different smoothing functions, which are shown in Fig.\ref{fig:smoothing}. The smoothing parameters $\mathcal{P}$ and $\mathcal{Q}$ are used to control how quickly the curve changes from one arc segment to another, which in turn determines the steepness of the slopes in the final driving pulse. This particular set of smoothing functions is of course by no means unique, and it may be possible to achieve slightly faster smooth pulses by modifying them. Regardless, we will see that this particular procedure is quite effective and can be straightforwardly applied to smooth all types of piecewise driving pulses. For simplicity, we will relax the constraint $|\Omega(t)|\leq 1$ here and instead fix the total gate time to be the same as the piecewise pulse we are starting from; the smoothing process generally leads to a slight modification of the maximal pulse amplitude, which can be corrected for at the end of the smoothing procedure by rescaling the pulse at the expense of a slight increase in its duration. 

\begin{figure}
\centering
\includegraphics[width=\columnwidth]{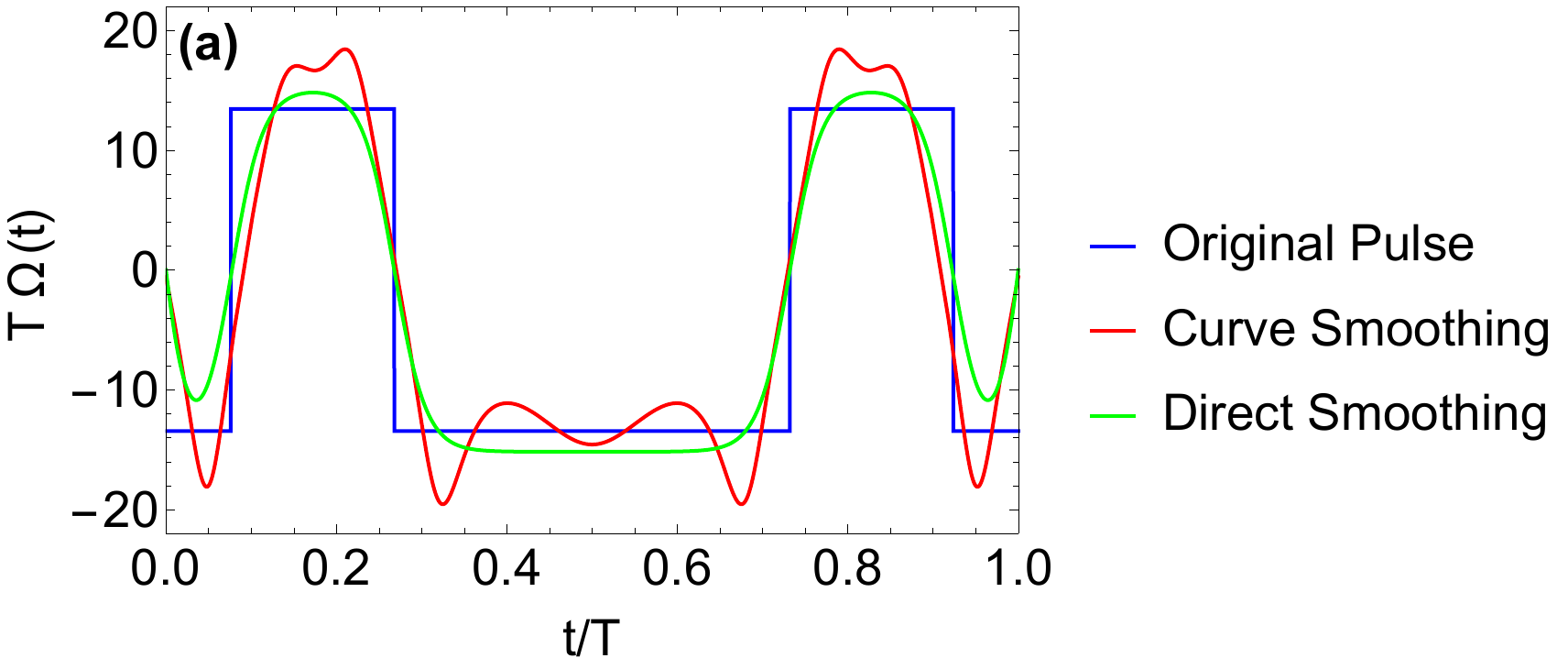}
\includegraphics[width=\columnwidth]{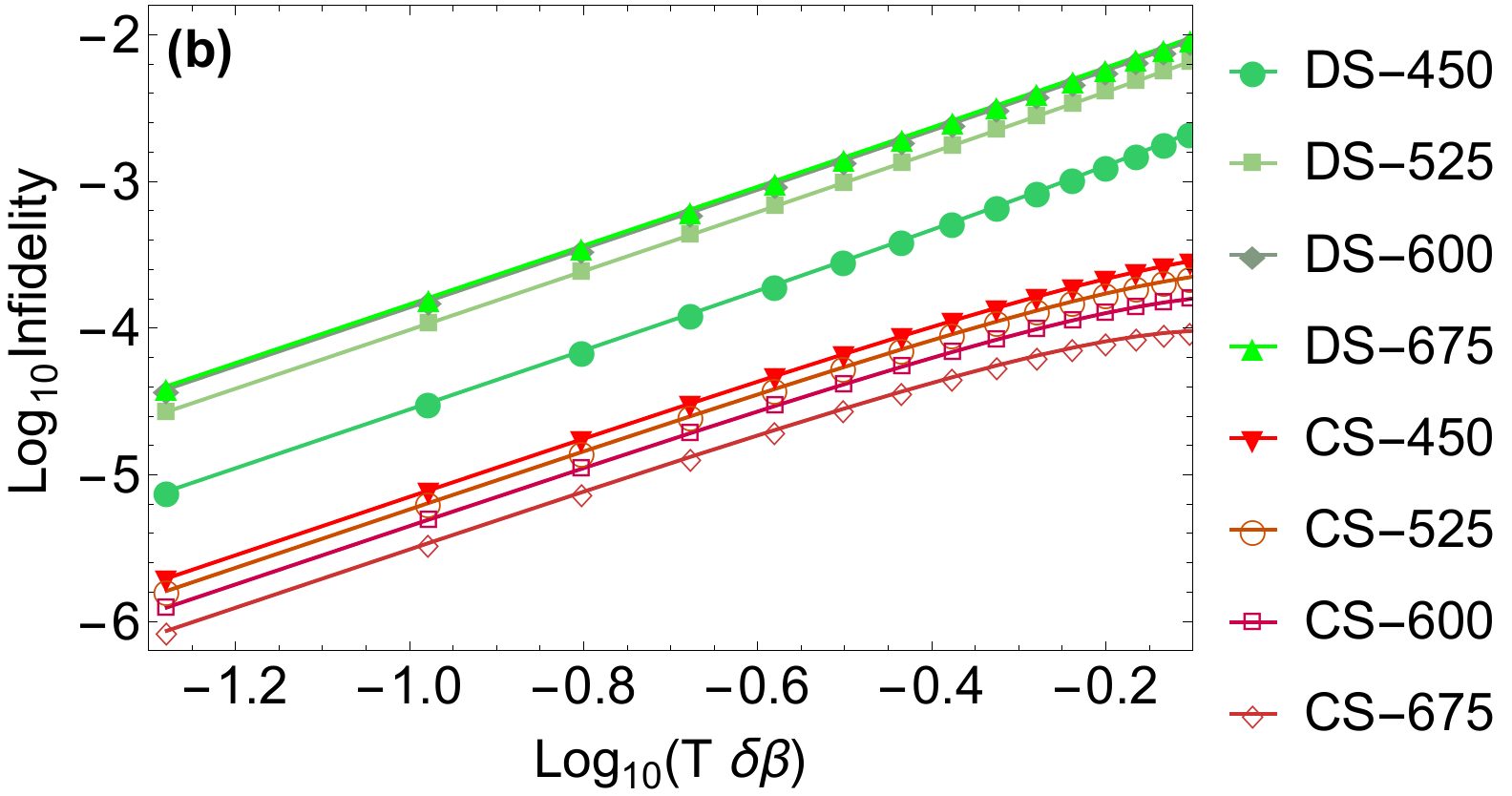}
\caption{Comparison of two types of pulse smoothing. (a) The original time-optimal composite square pulse with $\phi=0$ (blue), the pulse obtained by curve smoothing (red), and the pulse obtained from direct smoothing (green). (b) Fidelity versus noise strength for the two smoothed driving pulses shown in (a). Here \emph{CS} denotes curve smoothing and \emph{DS} direct smoothing. The numbers 450, 525, 600 and 675 are the maximum slopes of the driving pulses.}
\label{fig:comparison}
\end{figure}

Here, we focus on smoothing the second-order noise cancelling composite pulse given in Eq.~\eqref{eq:2ndorderpulse}. In this case, the functions $x(\lambda)$ and $y(\lambda)$ are each made up of five pieces $x_i(\lambda)$, $y_i(\lambda)$, with $1\leq i\leq 5$ corresponding to the five arcs in Fig. \ref{fig:secondorder}(a), with non-differentiable points $\lambda_i$ with $1\leq i \leq 4$, plus the origin point $\lambda_0=0$ and the final point $\lambda_f=\lambda(T)$. Explicit expressions for the $x_i$ for the case $\phi=0$ are given in Appendix~\ref{app:construction}. We show explicitly the smoothing procedure applied to $x(\lambda)$; the analogous steps should also be applied to $y(\lambda)$. The new, smoothed $x(\lambda)$ is constructed as follows:
\begin{equation}
\begin{split}
	 x(\lambda)=&f_1(\lambda_f-\lambda)\lambda x_1'(0)f_2(\lambda) \\+ &f_1(\lambda)\tilde{x}(\lambda)f_1(\lambda_f-\lambda)\\+&f_1(\lambda)(\lambda-\lambda_f)x_5'(\lambda_f)f_2(\lambda_f-\lambda).
	\end{split}
	\label{eq:smoothing}
\end{equation}
Here, the new $x(\lambda)$ function includes two linear functions, $\lambda x_1'(0)$ and $(\lambda-\lambda_f)x_5'(\lambda_f)$. By doing this, we guarantee that the second derivative of the function starts from zero and also ends at zero, as is required to ensure that the driving pulse starts and ends at zero. The function $\tilde{x}(\lambda)$ is defined as the smooth piecewise function
\begin{equation}
	\begin{split}
	\tilde{x}(\lambda)= &x_1(\lambda)f_4(\lambda-\lambda_1) +\\
	 &x_2(\lambda)f_3(\lambda-\lambda_1)f_4(\lambda-\lambda_2) +\\ 
	 &x_3(\lambda)f_3(\lambda-\lambda_2)f_4(\lambda-\lambda_3) + \\
	 &x_4(\lambda)f_3(\lambda-\lambda_3)f_4(\lambda-\lambda_4)+\\
	 &x_5(\lambda)f_3(\lambda-\lambda_4).
	\end{split}
\end{equation}
The fact that $x(\lambda)$ (and similarly $y(\lambda)$) is now fully analytical follows from the fact that all four smoothing functions are infinitely differentiable, and it guarantees that the new driving pulse is smooth. After obtaining $x(\lambda)$, $y(\lambda)$, it is necessary to recompute $t(\lambda)$ using Eq.~\eqref{eq:time} before applying Eq.~\eqref{eq:kappa} to find $\Omega(t)$. It should be noted that after applying the smoothing procedure, the final driving pulse will generally implement a rotation that deviates slightly from the original target rotation, but this can be corrected by rescaling the pulse appropriately. The net enclosed area of the curve may also deviate slightly from zero after the smoothing, meaning that the second order-error will not completely cancel. However, this deviation proves to be very minor in practice as can be seen from numerical simulations. In Fig.\ref{fig:comparison}(a), we compare the driving pulse obtained from our ``curve smoothing'' procedure with a smooth pulse obtained by performing a ``direct smoothing'' of the original square sequence in which each jump is replaced by a smoothly interpolating function of the form $f(t)=\tanh(\mathcal{R}(t-t_0))$. The two smoothing procedures are designed to respect the same upper bound on the pulse rise time. The original pulse is also shown and corresponds to the time-optimal sequence that implements a second-order-robust $\pi$-rotation.
Fig.~\ref{fig:comparison}(b) compares the $\pi$-rotation fidelities as a function of noise strength for both the curve smoothing and direct smoothing pulses for several different rise time constraints. It is evident that the curve smoothing pulses systematically outperform the direct smoothing ones by at least an order of magnitude regardless of the pulse rise times. Attempts to directly implement the idealized original pulse in experiments likely lead to actual pulses that are similar to the direct smoothing one depicted in Fig.~\ref{fig:comparison}(a) in the sense that they approximate the ideal pulse in a way that does not respect the error-cancellation constraints. This in turn leads to an incomplete cancellation of noise errors and to poorer performance. Our results illustrate that such errors can be avoided by treating error-cancellation and rise time constraints at the same footing.

\section{Conclusions}\label{sec:conclusions}

In summary, we exploited a recently discovered geometrical framework to reformulate the problem of finding the fastest possible error-suppressing pulses as a geometrical optimization problem. We solved this optimization problem to find the fastest pulses that implement single-qubit operations while suppressing noise to second order. Because these time-optimal driving fields are composite square pulses, we developed a smoothing algorithm to systematically obtain experimentally feasible pulses that respect rise time constraints while remaining fast and robust to noise. We showed that such pulses can reduce the infidelity by an order of magnitude compared to naively implementing ideal square pulses in the presence of rise time constraints. Our results show that the geometrical framework for dynamical gate correction provides simple and general ways to generate optimized driving pulses in the presence of experimental constraints, offering a powerful tool in the design of robust single-qubit operations.

\section*{Acknowledgments}

This work is supported by the Army Research Office (W911NF-17-0287) and by the Office of Naval Research (N00014-17-1-2971).

\appendix

\section{Derivation of the geometrical framework}\label{app:derive}

For a Hamiltonian given as in Eq. \eqref{eq:hamil}, we can write down the associated evolution operator parameterized by two time-dependent complex functions $u_1(t)$ and $u_2(t)$:
\begin{equation}
\mathcal{U}(t)=\left(
\begin{array}{cc}
u_1(t) & -u_2^*(t)\\
u_2(t) & u_1^*(t)\\
\end{array}
\right).
\end{equation}
For a general driving field $\Omega(t)$, it is not possible to obtain $u_1(t)$ and $u_2(t)$ in closed-form by solving Schr\"odinger's equation $i\dot{\mathcal{U}}(t)=\mathcal{H}(t)\mathcal{U}(t)$. However, we can instead obtain solutions as a power series in $\delta\beta$:
\begin{equation}
	\begin{split}
	&u_1(t)=e^{-i \theta(t)/2} (g_0(t)-g_2(t)\delta\beta^2+g_4(t)\delta\beta^4-...),\\
	&u_2(t)=-ie^{i\theta(t)/2}(g_1^*(t)\delta\beta-g_3^*(t)\delta\beta^3+...),\\
	\end{split}
\end{equation}
where $\theta(t)=\int_0^t\Omega(\tau)d\tau$ is the qubit rotation angle. The coefficient functions $g_i(t)$ obey a recurrence relation:
\begin{equation}
	g_i(t)=\int_0^te^{-i\theta(\tau)}g_{i-1}^*(\tau)d \tau,\label{eq:recurrence}
\end{equation}
with $g_0(t)=1$. Since these coefficient functions are complex, we can sketch the first-order function as a 2D curve:
\begin{equation}
	g_1(t)=e^{i \psi}(x(t)+iy(t)),\label{eq:g1}
\end{equation}
where $\psi$ is a global phase that can be chosen arbitrarily. The functions $x(t)$ and $y(t)$ are related  through the condition
\begin{equation}
 	\dot{x}(t)^2+\dot{y}(t)^2=1.
 \end{equation}
Using Eqs.~\eqref{eq:recurrence} and \eqref{eq:g1}, the following expression for the driving field can be derived:
\begin{equation}
 	\Omega(t)=\frac{\dot{x}\ddot{y}-\dot{y}\ddot{x}}{(\dot{x}^2+\dot{y}^2)^{3/2}},
 \end{equation}
which has the geometrical meaning that the driving pulse at any particular point along the curve equals the curvature at that point. In addition, it can also be shown that time measures arc length along the curve:
\begin{equation}
 	t(\lambda)=\int_0^\lambda\sqrt{x'(\mu)^2+y'(\mu)^2}d\mu.
 \end{equation}

In order to make a qubit rotation robust to first order, we need to impose $g_1(T)=0$ (where $T$ is the final time), implying that $x(T)=y(T)=0$ must be satisfied. This gives rise to the geometrical picture in which first-order robust driving pulses map to closed loops in the 2D plane.

Assuming that the first-order error-cancellation requirement is satisfied, the second-order error-cancellation requirement becomes
\begin{equation}
 \begin{split}
 	g_2(T)&=\int_0^T g_1'(\tau)g_{1}^*(\tau)d \tau \\
 	&=i\int_0^T\left(y'(\tau)x(\tau)-x'(\tau)y(\tau)\right)d\tau\\
 	&=2iA,
\end{split}
\end{equation}
where A is the area enclosed by the 2D plane curve. Thus, canceling the second-order noise error is equivalent to making the enclosed area zero.

\section{Optimal closed loop construction}\label{app:construction}

\begin{figure}
\centering
\includegraphics[width=\columnwidth]{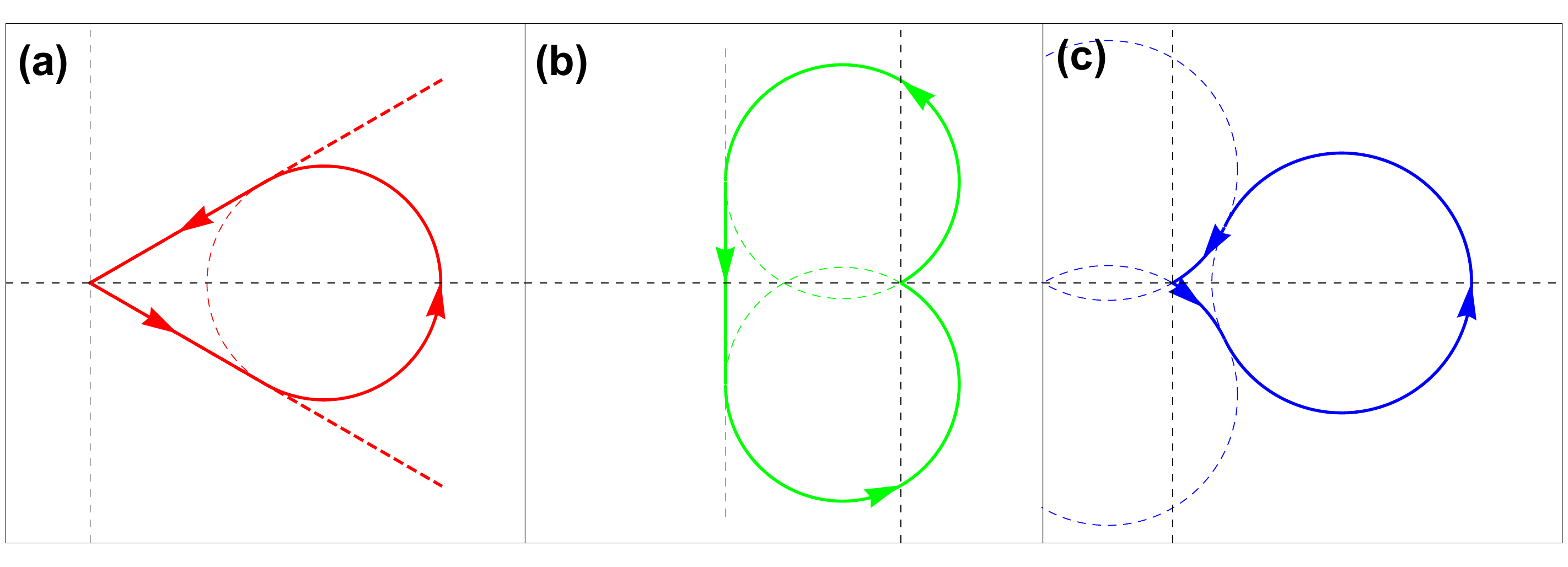}
\includegraphics[width=0.6\columnwidth]{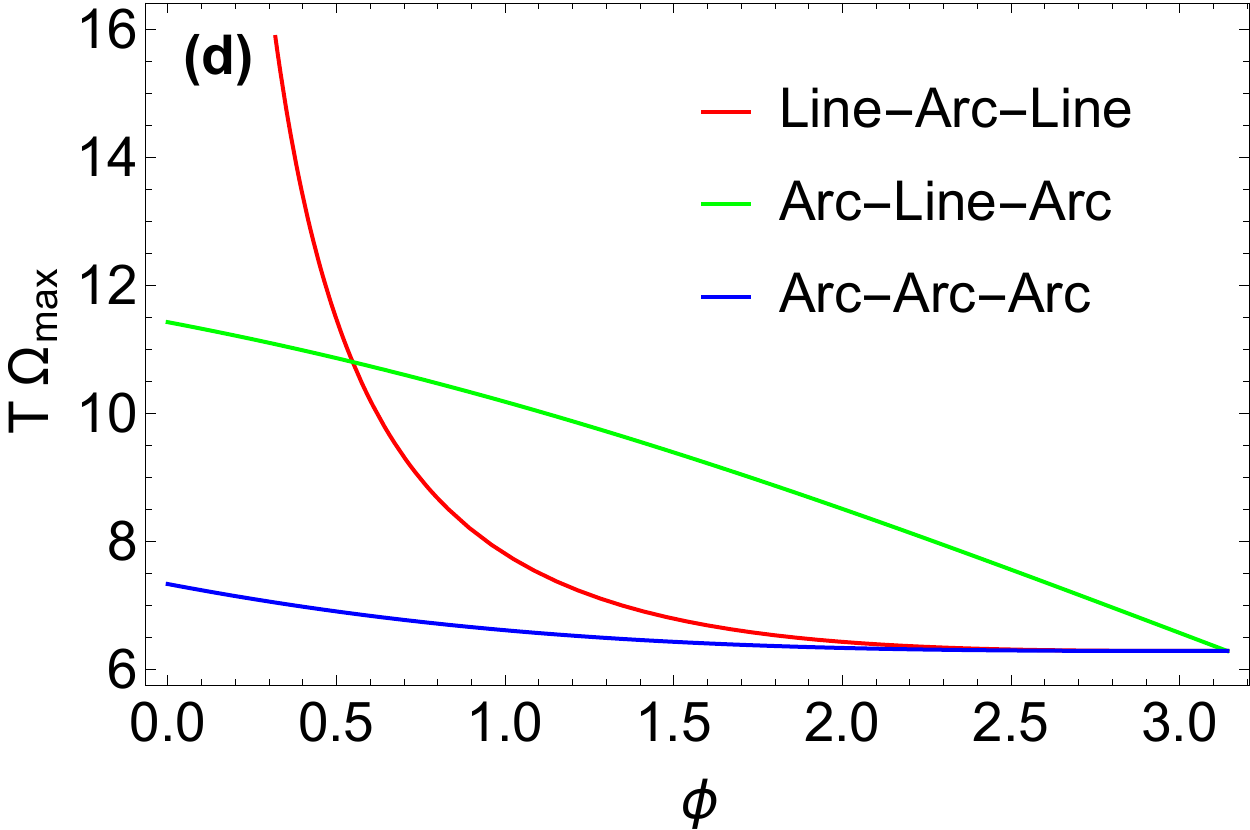}
\caption{Three piecewise curves that give time-optimal pulses. (a) An arc tangential to two straight lines. The length of such a curve is $\pi+\phi+2\cot(\frac{\phi}{2})$. (b) Two arcs tangential to one vertical line. The length is $3\pi-\phi+2\cot(\frac{\phi}{2})$. (c) Three arcs tangential to each other. The length is $4\psi+\pi-\phi$, where $\psi=\arccos(\cos(\phi/2)/2)$. (d) Comparison of the lengths of the three different curves. The curve shown in (c) is always the shortest.}
\label{fig:firstordercurves}
\end{figure}

In the main text, we show that a time-optimized control pulse canceling the first-order error maps to a curve made up of only two ingredients: straight lines, and $r=1$ arcs. We also argue that only three symmetric curves with three segments each are possible in the case of a general cusp angle at the origin. More complicated curves exist, but they will necessarily be longer and thus lead to longer gates. The three shortest curves combine the two basic building blocks in the following configurations: (1)\emph{Line-Arc-Line}, (2) \emph{Arc-Line-Arc} and (3) \emph{Arc-Arc-Arc}. Examples of such curves are shown in Fig.\ref{fig:firstordercurves}(a)-(c). 

The length of these curves can be easily calculated as functions of the subtended angle $\phi$. As shown in Fig.\ref{fig:firstordercurves}(d) we find that the third case, where the curve is comprised of three tangential arcs, is much shorter than the other two cases. We thus conclude that the curve shown in Fig.\ref{fig:firstordercurves}(c) is the globally optimal solution and thus yields the fastest possible first-order dynamically corrected gates.

\begin{figure}
\centering
\includegraphics[width=\columnwidth]{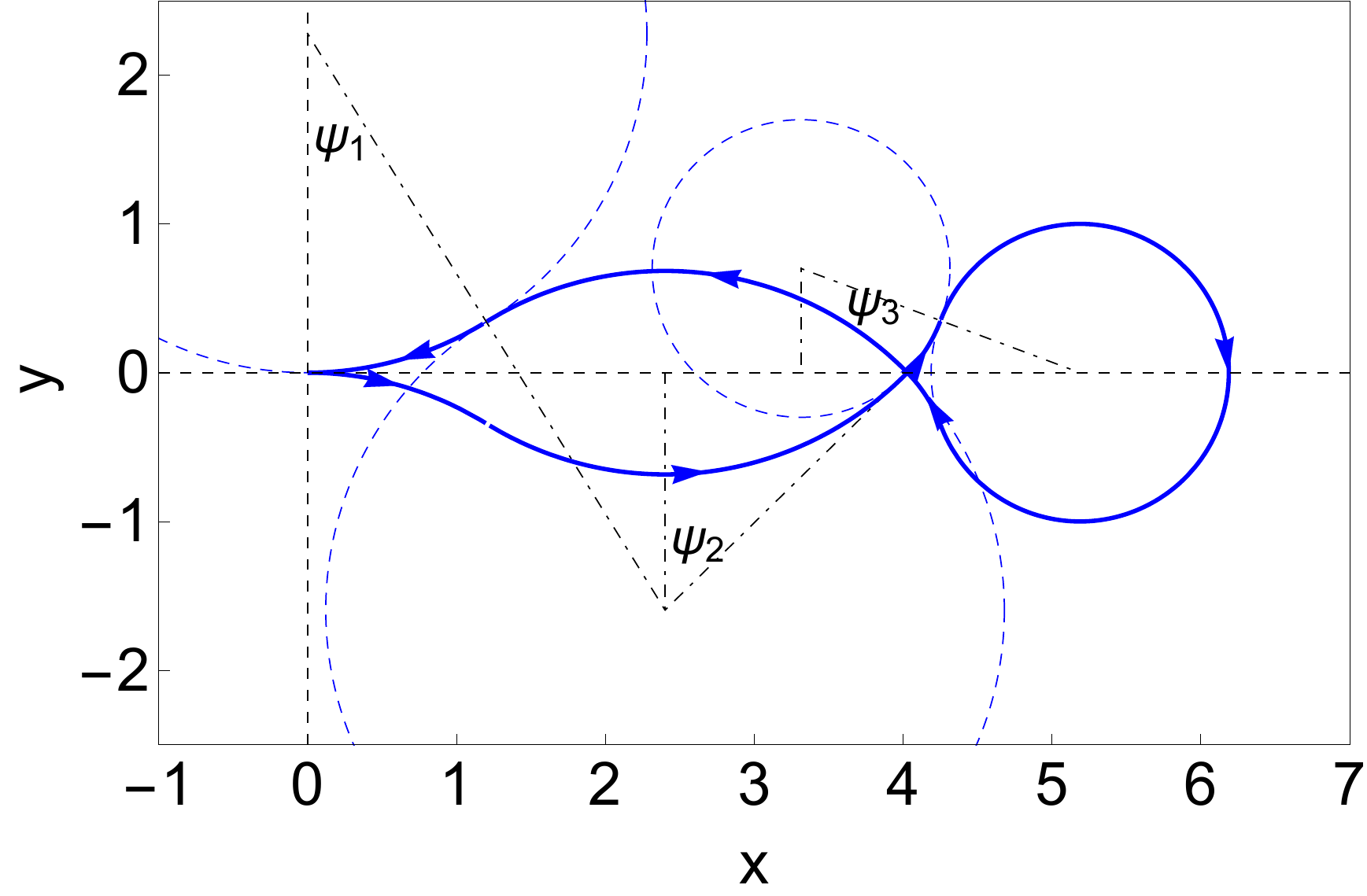}
\caption{An example of a second-order error-canceling curve made up of externally tangential arcs.}
\label{fig:secondordercurves}
\end{figure}

For second-order error cancellation, we should have at least one crossing in the curve, in which case the curve will contain two loops. Given that in the first-order analysis presented above we found that curves made up of $r=1$ arcs have very short lengths for a given cusp angle, we focus our search on curves with a similar structure. In particular, we consider an arrangement in which the loop furthest from the origin forms a closed curve that is essentially identical to the first-order error-cancellation case, and in which the loop nearer to the origin also has a similar shape except that we allow for a kink at the point where two loops touch, and the radius is allowed to be greater than one. An example of such a curve is shown in Fig.~\ref{fig:secondordercurves}. Here, in order to illustrate how the parameters are defined, we deliberately did not optimize them as we did in Fig.~\ref{fig:secondorder}(a), and we instead fixed $k=0.7$. For this type of curve, we find that the total net area is given by
\begin{widetext}
\begin{equation}
	A=\frac{r^2 \left(\tan (\psi_1) \left(-2 k^2+\cos (\phi )+1\right)-2 \sqrt{1-k^2} k+2 \psi_2+\phi -\sin (\phi )\right)+\sec (\psi_3) \sin (2 \psi_2-\psi_3)-2 \psi_2-\tan (\psi_3)-\pi }{2},
\end{equation}
\end{widetext}
where $\psi_1=\arccos(\frac{\cos(\phi/2)+k}{2})$, $\psi_2 = \arccos(k)$ and $\psi_3=\arccos(k/2)$. Here, $rk$ is the distance between the center of the second arc and the $x$-axis. We can easily obtain the solution $r(k,\phi)$ using the fact that the net area vanishes, and it turns out that for each $\phi$, $r(k)$ monotonically increases with k. In addition, the perimeter of this curve is given by
\begin{equation}
	r (4 \psi_1+2 \psi_2-\phi )-2 \psi_2+4 \psi_3+\pi.
\end{equation}
In order to minimize this length, we find that it is necessary to minimize $k$, which is equivalent to searching for the lowest possible value of $r$. Here, the lowest $r$ would be the lowest allowed radius: $r=1$. Plugging this back into the zero-net-area constraint, we can obtain the numerical solution for $k(\phi)$, as shown in Fig.~\ref{fig:secondorder}(c).

Once the optimized parameters are applied, the parametric function defining the curve becomes a 5-part piecewise function with $1\leq i\leq 5$:
\begin{equation}
\begin{aligned}[c]
	x(t)=x_i(t),\\
	y(t)=y_i(t), 
	\end{aligned}
	\qquad
	\begin{aligned}[c]
	t_{i-1}<t\leq t_i,\\
	t_{i-1}<t\leq t_i.\\
	\end{aligned}
\end{equation}

To provide a concrete example, we consider the case where the subtended angle is $\phi=0$, which corresponds to a $\pi$-rotation. In this case, we have
\begin{widetext}
\begin{equation}
\begin{aligned}[c]
	x_1(t)&= \sin (t) \\
 x_2(t)& = \cos (3.62163\, -t)+1.70986  \\
 x_3(t)& = \cos (6.72573\, -t)+3.70951  \\
 x_4(t)& = \cos (9.82983\, -t)+1.70986 \\
 x_5(t)& = \cos (11.8807\, -t)
\end{aligned}
\qquad
\begin{aligned}[c]
 y_1(t)&= \cos (t)-1 \\
 y_2(t)& = 0.0374882\, -\sin (3.62163\, -t)  \\
 y_3(t)& = \sin (6.72573\, -t) \\
 y_4(t)& = -\sin (9.82983\, -t)-0.0374882  \\
 y_5(t)& = \sin (11.8807\, -t)+1 
\end{aligned}
\qquad
\begin{aligned}[c]
t_1 & = 1.02542 \\
t_2 & = 3.60288 \\
t_3 & = 9.84858 \\
t_4 & = 12.426 \\
t_5 & = 13.4515 
\end{aligned}
\label{eq:curvedetail}
\end{equation}
\end{widetext}

We can obtain smoothed versions of $x(\lambda)$ and $y(\lambda)$ by plugging into Eq.~\eqref{eq:smoothing}. We can also obtain $t(\lambda)$ using Eq.~\eqref{eq:time}. Thus by interpolating $\{\Omega(\lambda), t(\lambda)\}$, we can obtain our final smoothed function $\Omega(t)$ which approximates the original composite square pulse in an optimal way.

\bibliographystyle{apsrev}
\bibliography{note}

\end{document}